\documentclass[aps,prl,reprint,groupedaddress]{revtex4-1}

\usepackage{amsmath}
\usepackage{graphicx}
\usepackage{epstopdf}

\newcommand{\ket}[1]{\mbox{$|#1\rangle$}}

\begin{document}

\title{Quantum Entanglement of Complex Photon Polarization Patterns in Vector Beams}

\author{Robert Fickler$^{1,2}$}
\author{Radek Lapkiewicz$^{1,2}$}
\author{Sven Ramelow$^{1,2,3}$}
\author{Anton Zeilinger$^{1,2}$}
\affiliation{$^1$Vienna Center for Quantum Science and Technology, Faculty of Physics, University of Vienna,
Boltzmanngasse 5, Vienna A-1090, Austria
\\
$^2$Institute for Quantum Optics and Quantum Information, Austrian Academy of Sciences,
Boltzmanngasse 3, Vienna A-1090, Austria
\\
$^3$current address: Cornell University, 271 Clark Hall, 142 Science Dr., Ithaca, 14853 NY, USA
}

\begin{abstract}
We report the efficient creation and detection of hybrid entanglement between one photon's polarization and another photon's complex transverse polarization pattern. The polarization measurement of the first photon triggers a polarization sensitive imaging of its partner photon, the vector photon, using a single-photon sensitive camera. Thereby, we reconstruct tomographically the vector photon's complex polarization patterns dependent on the type of polarization measurement performed on its partner. We visualize the varying strengths of polarization entanglement for different transverse regions and demonstrate a novel feature: each vector photon can be both entangled and not entangled in polarization with its partner photon. We give an intuitive, information theoretical explanation for our results.
\end{abstract}

\date{\today}

\maketitle

Polarization is one the most prominent degrees of freedom (DOF) for photonic quantum information and already relies on mature technologies for creation, manipulation and analysis of quantum features \cite{zeilinger2005happy}. Quantum information encoded in transverse spatial modes of photons, e.g. Laguerre Gauss (LG) modes, has attracted increasing attention \cite{molina2007twisted} due to interesting quantum phenomena like quantized orbital angular momentum \cite{allen1992orbital, mair2001entanglement} and higher dimensional entanglement \cite{vaziri2002experimental, molina2005experimental, dada2011experimental, krenn2013studies}. The non-trivial combination of both DOFs, more precisely the superposition of two different, orthogonally polarized spatial modes results in the so-called vector-polarization beams. Their common feature is a transversely varying polarization. Of these, ``cylindrical vector beams" show cylindrical symmetry in polarization \cite{zhan2009cylindrical}. They have advantages for applications like improved excitation of plasmons \cite{kano1998excitation}, increased coupling to individual atoms \cite{sondermann2007design} or sharper focusing \cite{dorn2003sharper}. Another class are the so-called ``Poincar\'e Beams" which contain every polarization on the Poincar\'e sphere \cite{beckley2010full} and show interesting features, like various types of polarization singularities \cite{nye1983lines, freund2002elliptic, dennis2002polarization, galvez2012poincare} or changes of the polarization pattern, while freely propagating \cite{cardano2013generation}. Albeit offering a great possibility to explore rich optical patterns, vector beams are nearly unexplored experimentally in the quantum regime. One recent demonstration used spin-orbit hyperentangled photon pairs from a down-conversion process, i.e. pairs that are entangled in polarization and the transverse spatial mode, to remotely prepare vector-polarization states \cite{barreiro2010remote}. However, phase matching in the down conversion process, constrains such an approach to a few possible vector-polarization states.
\begin{figure}
\centering  \includegraphics[width=0.48\textwidth ]{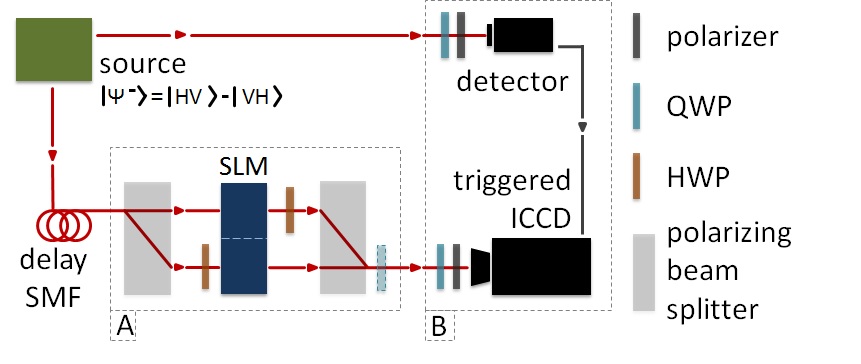}
\caption{Sketch of the experimental setup. Polarization entanglement between two photons is created (green box) in a down conversion process. One photon is coupled to a single-mode fiber (SMF), delayed by 35m and brought to the interferometric transfer setup (A). Here, the path is coherently split according to polarization and the photon state is transferred by a spatial light modulator (SLM) to a different transverse spatial mode for each path. The vector-polarization state results by recombining the two modes. An removable quarter-wave plate (QWP) rotates the linear polarization components to circular ones in case of LG and Poincar\'e vector photon creation. (B) To demonstrate the generated hybrid entanglement a polarization sensitive coincidence imaging is performed. The polarization measurement of the unchanged photon (QWP, polarizer, single photon detector) is used to trigger an intensified CCD camera (ICCD). For each trigger polarization a spatially resolved polarization tomography of the vector photon is performed (QWP, polarizer, ICCD) and used to reconstruct its complex polarization pattern. \label{fig1}}
\end{figure}

In this letter, we present an experiment where we create and study photon pairs with hybrid entanglement between polarization and a rich variety of different vector-polarization states. We test polarization entanglement at local transverse positions with three different entanglement criteria, thereby visualizing experimentally the different strength of these tests. Moreover, we demonstrate a novel feature of entangled vector beams: the hybrid bi-photon state can be either entangled or separable in polarization -- dependent on the transverse spatial position. In addition, we demonstrate the flexibility of our method by creating custom-tailored polarization patterns which also contain every polarization, like Poincar\'e modes.

\begin{figure}
\centering  \includegraphics[width=0.45\textwidth ]{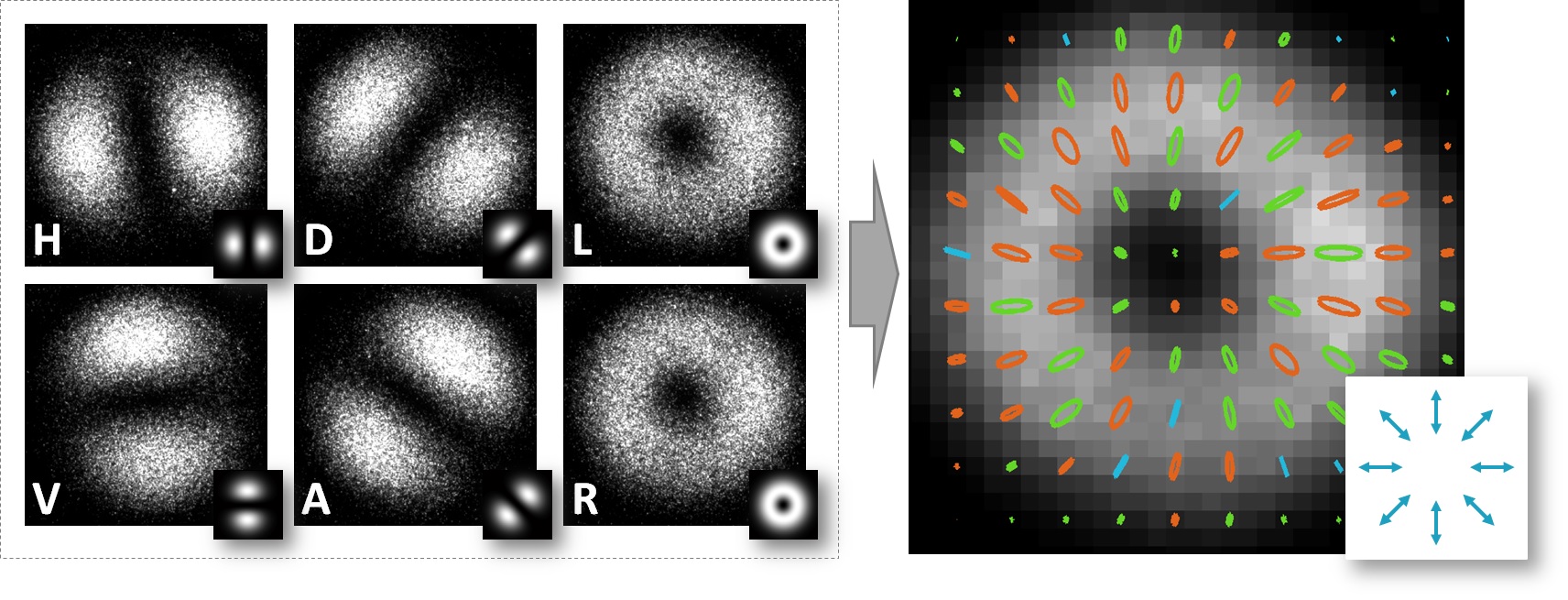}%
\caption{Example of a transverse polarization tomography from coincidence images. The presented measurements were done with diagonally polarized trigger photons and vector photons composed of circularly polarized, first order LG modes $l$=+1 and $l$=-1. Left: The recorded intensity distribution (accumulating single photons for 15 seconds) depends on the polarization settings in front of the ICCD camera (white letters; H, V, D, A, R, L for horizontal,vertical,diagonal,anti-diagonal,left and right hand circular respectively). Right: Local polarization tomography (10x10 pixel arrays) results in the reconstructed pattern of a cylindrical vector photon, more precisely a radially polarized state. The colors depict the type and handedness of polarization (blue=linear, red=right elliptic, green=left elliptic). The average intensity distribution is overlaid with the reconstructed polarization pattern. Both, the recorded coincidence images as well as the reconstructed pattern fit well to the theoretical predictions (small insets). \label{fig2}}
\end{figure}

In our setup (Fig. \ref{fig1}), we first create polarization entangled photon pairs with a parametric down conversion process in a Sagnac configuration (405nm cw pump laser, 20mW pump power, 810nm downconverted photon pairs, approx. 500kHz coincidence counts)  \cite{kim2006phase, fedrizzi2007wavelength}. One photon is unchanged while the other is coupled to a single-mode fiber for Gauss mode filtering, delayed by 35m and brought to a second setup. There, the photon is coherently converted into a vector-polarization mode (herein called ``vector photon") by interferometrically superposing two different spatial modes with orthogonal polarization \cite{tidwell1990generating}. The resulting hybrid-entangled two-photon state can be written
\begin{equation}
\ket{\psi}=a\ket{H}\ket{spM_{H/R}}+e^{i\phi}b\ket{V}\ket{spM_{V/L}} \label{eq_state}
\end{equation}
where $a$, $b$, and $\phi$ are real and $a^2 + b^2 = 1$; $H$ and $V$ denote the horizontal and vertical polarization of the unchanged photon; $spM$ and its index represents the different spatial modes and their polarizations (H, V, R, L for horizontal, vertical, right and left hand circular respectively); the positions of the ket-vectors label the different photons. Due to the flexibility of the SLM, a huge variety of different vector beams including ``cylindrical vector beams", showing cylindrical symmetry in their polarization pattern \cite{zhan2009cylindrical}, and ``Poincar\'e beams", containing all polarizations on the Poincar\'e sphere \cite{beckley2010full}, can be realized \cite{maurer2007tailoring, galvez2012poincare}. The crucial phase-stability of the interferometer was assured by a folded Sagnac-like structure \cite{fickler2012quantum}. The polarization pattern of the vector photon depends on the type and result of the polarization measurement of its partner photon. Therefore, we performed a coincidence imaging measurement \cite{fickler2013real} extending it with polarization analysis. The single-photon detector signal of a polarization measurement is used as a trigger for an intensified CCD camera (ICCD) (Andor iStar A-DH334T-18U-73, 5ns coincidence window, ~20\% quantum efficiency, effective pixel size 13x13$\mu$m). From a polarization tomography of the vector photon, performed by a polarizing filter and the ICCD camera, the complex polarization pattern can be reconstructed with very high spatial resolution (Fig. \ref{fig2}).

\begin{figure}
\centering  \includegraphics[width=0.48\textwidth ]{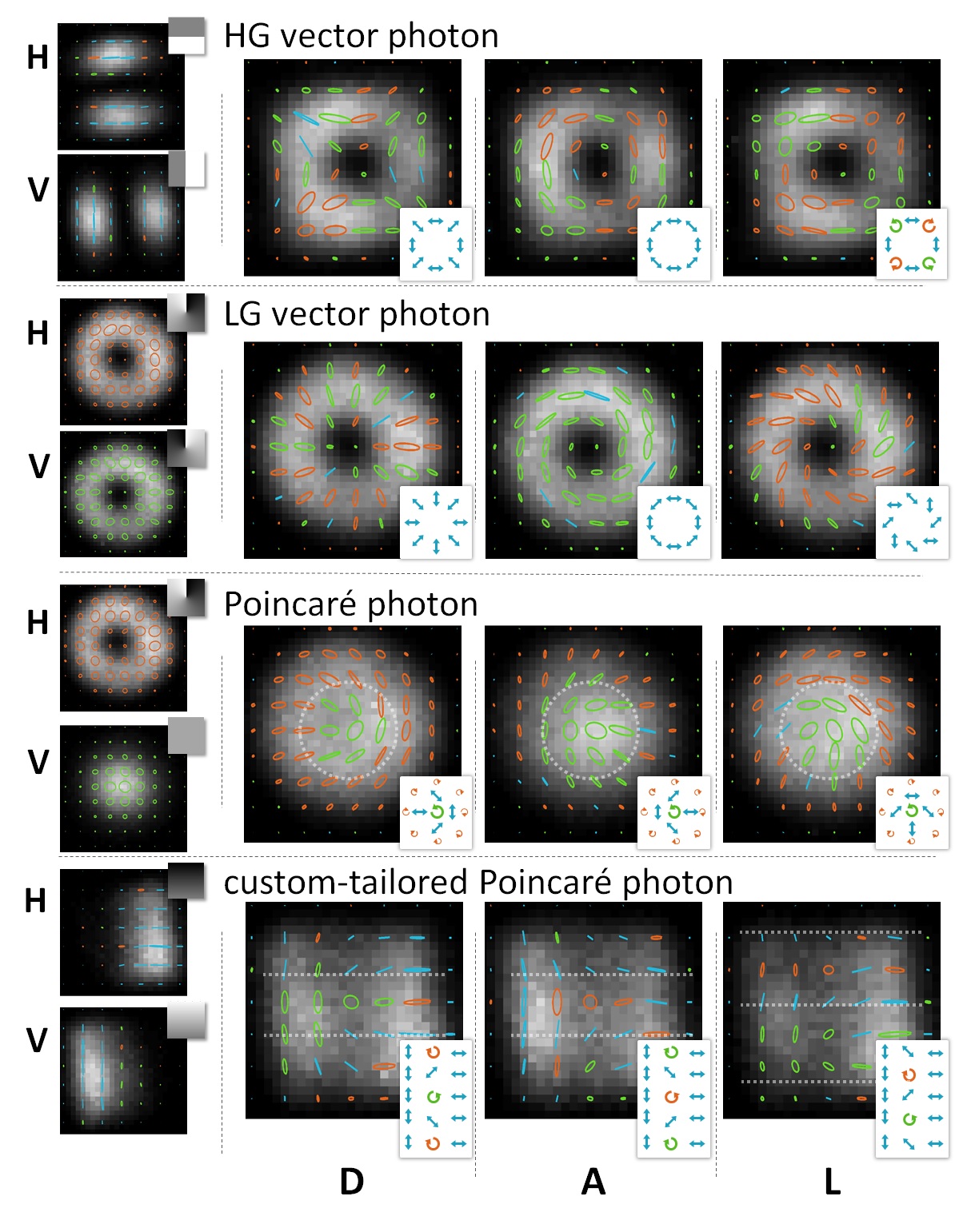}%
\caption{Polarization patterns measured for different vector photons entangled with polarized partner photons. The average registered intensity distribution for each trigger polarization (black letters) is shown and overlaid with the measured polarization pattern (every 4\textsuperscript{th} polarization shown). Colors depict the type and handedness of the measured polarization (coding as in Fig. \ref{fig2}). For H or V polarized trigger photons (first column) the polarization is uniformly distributed over the spatial mode. The imprinted phases are presented by gray scale insets (upper right; linearly from 0 = black to 2$\pi$ = white). If the trigger photon is found in a superposition of H and V (columns two to three), the vector photon shows different polarization patterns (theoretical pattern depicted in insets on the lower right). R polarized triggers are omitted, because they are only mirrored patterns of L polarized triggers. Cylindrical vector beam patterns, including radial, azimuthal and spiral configurations, can be found for HG and LG vector photons (first two rows). In the third row, patterns of Poincar\'e photons are shown which exhibit a C-point singularity in the center and a circular L-line singularity around it (dashed circle). Similar C-points and L-lines can be found in patterns of the custom-tailored Poincar\'e photon (last row; dashed line). \label{fig3}}
\end{figure}

Due to entanglement, different polarizations of the trigger photon result in different polarization patterns of its partner vector photon, thus the Bloch sphere for vector photons --- the higher order Poincar\'e sphere --- can be imaged \cite{milione2011higher}\cite{holleczek2010classical}. The polarization patterns of cylindrical vector photons built by linearly polarized HG modes or circularly polarized LG modes can be distinctly recognized for different trigger polarizations (Fig. \ref{fig3}, first two rows). In addition, Poincar\'e photons are remotely generated which exhibit various polarization singularities, like C-points (orientation of the polarization ellipse is undefined) or L-lines (handedness of the polarization is undefined) \cite{dennis2002polarization, galvez2012poincare} (Fig. \ref{fig3}, third row). If the SLM surface is imaged on the ICCD camera chip and the diffraction efficiency of the displayed hologram is adjusted according to the desired intensity structure, the phase and the intensity of the photons can be modulated \cite{leach2005vortex}. With this technique it is possible to create and entangle any custom-tailored polarization pattern for the vector photon. We demonstrate this remarkable flexibility by creating a square shaped vector beam consisting of two linearly polarized modes (Fig. \ref{fig3}, fourth row). For each mode the intensity changes linearly from left to right and the phase varies linearly from top to bottom. This results in a polarization pattern with a continuous change from V (left side) to H (right side) through all possible polarizations, as a function of the vertical position.  

\begin{figure}
\centering  \includegraphics[width=0.42\textwidth ]{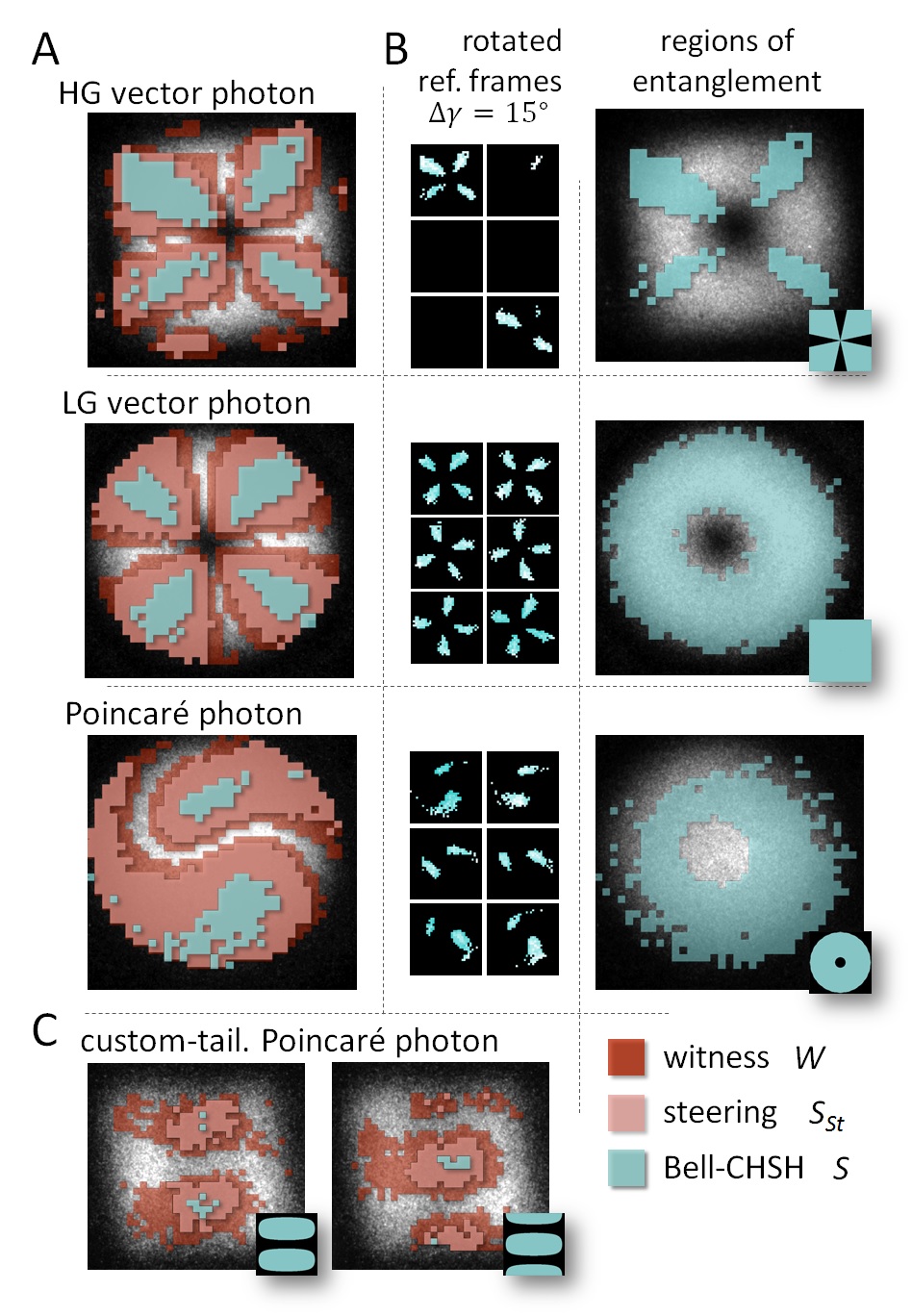}%
\caption{Polarization entanglement of the created hybrid entanglement bi-photon state. (A) Three different tests for polarization entanglement were performed: witness (dark red, Eqs. \ref{eq:witness}), steering (light red, Eqs. \ref{eq:steering}), Bell-CHSH (blue, Eqs. \ref{eq:bell}). For LG vector photons and Poincar\'e photons the Bell-CHSH inequality was tested with L, R, D and A polarization triggers instead of the usual H, V, D and A polarization settings. The different strengths of the criteria are directly visible in the size of the non-classical regions, i.e. regions where more than 100 photons contributed to every term of each criterion and the classical bound is exceeded by 3 standard deviations (Poissonian count statistics assumed). (B) If the polarization setting in front of the ICCD camera is altered (rotation of the reference frame) to account for the varying polarizations within the vector photon's polarization pattern an interesting entanglement pattern appears. Only where the modal overlap of the two components is big enough (and enough photons were detected) can entanglement in polarization be revealed (small insets: theory). (C) Similarly for custom-tailored Poincar\'e photons, the modal overlap is too small on the left and right parts of the beam irrespectively of the trigger polarization settings (left and right measurement and theory insets) thus regions of entanglement are rarely found there. \label{fig4}}
\end{figure}

So far, the high-contrast intensity images for different trigger polarizations and the subsequent changes of the polarization patterns have only suggested the successful generation of entanglement. However, to demonstrate the non-classicality of the state quantitatively, we evaluate locally ``for every hybrid entangled photon pair" three different types of entanglement measures in the polarization DOF (Fig. \ref{fig4}). We register coincidence images for appropriate polarization combinations (trigger and image polarization) and evaluate the average photon number within regions of 10x10 pixels. From these local measurements we calculate the value for each entanglement criterion. Note, that this is only feasible with our real-time coincidence imaging technique \cite{fickler2013real}. Relying on counting single-photon events in sparse images or scanning single-pixel detectors across the beam would have been extremely challenging and time consuming. The first measure of entanglement is an entanglement witness fully relying on quantum mechanical predictions \cite{guhne2009entanglement}
\begin{equation}
W=|\sigma_x \otimes \sigma_x |+|\sigma_y \otimes \sigma_y |+|\sigma_z \otimes \sigma_z | \leq 1 \label{eq:witness}
\end{equation}
where $\sigma_x$, $\sigma_y$ and $\sigma_z$ stand for the single-qubit Pauli matrices (mutually unbiased bases). The witness is bounded by 1 for any separable state, thus values larger than 1 verify entanglement. As a second criterion we used a steering inequality \cite{cavalcanti2009experimental}
\begin{equation}
S_{St}=|\sigma_x \otimes \sigma_x |^2+|\sigma_y \otimes \sigma_y |^2+|\sigma_z \otimes \sigma_z |^2 \leq 1 \label{eq:steering}
\end{equation}
If measurement results exceed the bound, the non-classicality of the state i.e. ``non-local steering of the vector photon" is proven. Because weaker assumptions are made than for the witness (Eqs. \ref{eq:witness}), namely only one side assumes quantum mechanics, the steering inequality is violated by a smaller class of states. In the last test for entanglement we use the Bell-CHSH-inequality \cite{bell1964einstein, clauser1969proposed}
\begin{equation}
S=|E(\alpha ,\beta )-E(\alpha' ,\beta )+E(\alpha ,\beta' )+E(\alpha' ,\beta' )| \leq 2 \label{eq:bell}
\end{equation}
where $\alpha$, $\alpha'$, $\beta$ and $\beta'$ denote different measurement settings (orientations of the polarizer) and E is the normalized expectation value for photon pairs to be found with these settings. A violation of this Bell-CHSH-bound proves entanglement of the created state without relying on any quantum mechanical assumptions, thus it excludes a larger class of states, i.e. every state described by local realism. In Fig. \ref{fig4} A, regions are shown where the witness, the steering or the Bell-CHSH inequalities prove entanglement. The measured results depict visually, that the weaker the assumptions of the criterion the smaller are the regions of successful entanglement demonstration.

Furthermore, a novel interesting feature of entangled vector photons can be demonstrated by testing locally for polarization entanglement (Fig. \ref{fig4} B). To account for the spatially varying polarization, different ``reference frames" i.e. polarizer angles can be used to measure entanglement at different regions of the transverse beam spread. Hereby, the substantial difference between the demonstrated vector photons becomes apparent: if they consist of two beams with the same spatial intensity profile (e.g. vector photons from circularly polarized LG modes of the same order), any transverse position shows polarization entanglement. In contrast, if vector photons are built by modes that have different transverse profiles, entanglement in the polarization DOF can only be found in regions where the overlap is big enough. Theoretically only very small regions are separable, because a small overlap between the two modes already leads to (non-maximally) entangled photons. To experimentally demonstrate the effect we used the Bell-CHSH-criterion (Eqs. \ref{eq:bell}) where the regions of proven entanglement are the smallest for each reference frame. This interesting feature of being entangled and separable in the polarization DOF at the same time can be explained intuitively: at transverse positions where a polarization measurement can reveal the path of the photon inside the interferometer (with which we created the vector photon) no superposition and thus no entanglement is measurable. 

In summary, we have used hybrid entanglement between polarization and different types of vector photons to demonstrate experimentally that entangled photons can exhibit complex polarization patterns including various polarization singularities. Furthermore, we have demonstrated, and given an information theoretic explanation for, a fundamentally interesting feature of the created complex bi-photon state: certain regions of the beam show different degrees of non-classicality in polarization while others do not at all. Apart from a better intuitive understanding of quantum mechanical features, we envision that our results will stimulate further investigations in quantum information experiments. It is known that the polarization pattern is essential for photon-atom-coupling efficiencies to be made close to 100\% \cite{sondermann2007design}. This suggests that entangled vector photons will facilitate entangling separated atoms in broader quantum networking tasks. Another interesting application might be cluster-state quantum computing where the polarization and spatial degree of freedom increases the amount of encoded qubits per photon \cite{andrews2012angular}. The generated states could be used for teleportation of complex structures and might be advantageous if information in several degrees of freedom needs to be encoded on a single photon. Furthermore, the novel features of these hybrid entangled states might improve quantum cryptographic schemes.

\begin{acknowledgments}
We would like to thank Mario Krenn for fruitful discussions. This work was supported by the European Research Council (advanced grant QIT4QAD, 227844) and the Austrian Science Fund (FWF) through the Special Research Program (SFB) Foundations and Applications of Quantum Science (FoQuS; Project No. F4006-N16) and the European Community Framework Programme 7 (SIQS, collaborative project, 600645). RF and RL are supported by the Vienna Doctoral Program on Complex Quantum Systems (CoQuS, W1210-2). SR is supported by a EU Marie-Curie Fellowship (PIOF-GA-2012-329851).
\end{acknowledgments}

\bibliographystyle{apsrev}
\bibliography{PRL_Fickler_BIB}{}

\end{document}